\documentclass[fleqn,usenatbib]{mnras}

\usepackage{newtxtext,newtxmath}

\usepackage[T1]{fontenc}

\DeclareRobustCommand{\VAN}[3]{#2}
\let\VANthebibliography\thebibliography
\def\thebibliography{\DeclareRobustCommand{\VAN}[3]{##3}\VANthebibliography}

\usepackage{graphicx}	
\usepackage{amsmath}	

\title[Variable star membership in globular clusters]{On the membership of variable stars in galactic globular clusters: The Oosterhoff gap}

\author[Prudil \& Arellano Ferro]{
Z. Prudil,$^{1}$\thanks{Equal first authors.}
A. Arellano Ferro,$^{2}$$^{\textcolor{blue}{\star}}$\thanks{Corresponding Author: E-mail: armando@astro.unam.mx}
\\
$^{1}$ European Southern Observatory, , Karl-Schwarzschild-Straße 2, 85748, Garching, Germany\\
$^{2}$Instituto de Astronom\'ia, Universidad Nacional Aut\'onoma de M\'exico, Ciudad de M\'exico, CP 04510, M\'exico.\\
}

\date{Accepted: ; Received: 2020  in original form 2020 July}

\pubyear{2020}

\begin{document}

\label{firstpage}
\pagerange{\pageref{firstpage}--\pageref{lastpage}}
\maketitle

\begin{abstract}
We have performed a critical evaluation of the membership status of all variable stars in globular clusters recorded in the Catalogue of Variable Stars in Globular Clusters (CVSGC) curated by Christine Clement. To this end, we employed the systematic and bulky membership analysis performed by E. Vasiliev and H. Baumgardt based on the proper motions and parallaxes given in \textit{Gaia}-EDR3. We found numerous variables in the CVSGC which are in fact field stars, which is particularly the case for globular clusters located in the Galactic bulge. Using the newly acquired list of reliable cluster members we examine the Oosterhoff dichotomy present among the Milky Way (MW) globular clusters using their RR~Lyrae stars content. We confirm the presence of the Oosterhoff gap, separating both Oosterhoff groups. The Oosterhoff gap is mostly populated by globular clusters associated with MW dwarf galaxies and globular clusters with a low number of fundamental mode RR~Lyrae variables. Several of the clusters in the Oosterhoff gap were previously linked to past merger events (e.g. Kraken/Heracles).
\end{abstract}

\begin{keywords}
globular clusters -- Horizontal branch -- RR Lyrae stars -- Fundamental parameters.
\end{keywords}


\maketitle

\section{Introduction}
\label{sec:intro}

Globular clusters (GC) harbor variable stars. The number of variables per cluster varies from zero to a couple of hundredths according to the Catalogue of Variable Stars in Globular Clusters\footnote{\url{https://www.astro.utoronto.ca/~cclement/cat/listngc.html}} curated by Prof. Christine Clement of the University of Toronto \citep{Clement2001}. The types of variables in GC are assorted, being the most common the RR Lyrae, semiregular red giants RGB/L, the SX Phe among the blue stragglers, the Pop II Cepheids, and eclipsing binaries. It has been known since early the XXth century that variable stars, particularly RR Lyrae and Cepheids, are of paramount importance in the determination of cosmic distances \citep{Shapley1917,Shapley1918,Baade1956,Sandage1970} and that their light curves morphology carries information of other physical parameters of astrophysical relevance \citep{vanAlbada1971}, such as stellar mass, luminosity, and metallicity.

When these indicators of physical parameters are in globular clusters, they can be used to estimate the average values of the parental cluster. Nevertheless, crowded fields in the sky may naturally include field stars that are projected against the field of a given cluster without having a physical connection and a common evolutionary nor chemical history. This is most common in clusters in the Galactic bulge where the field is extremely rich in variable stars, many of which are of the RR Lyrae type. Therefore, if estimating globular cluster properties from their variable star population is an aim, one must be able to distinguish true cluster members from spurious field stars.

In this work, we propose to assign a membership likely status to each one of the variables presently cataloged in the CVSGC, by making use of the \textit{Gaia}-EDR3 proper motions and the systematic membership analysis performed for 170 globular clusters by \citet{Vasiliev2021}. In the following Sections, we will describe in detail our procedures, point out the limitations and caveats, and highlight those clusters that for some reason or the other present peculiarities. In the end, we shall offer lengthy tables with our conclusion on the membership of the variables in a large sample of Galactic globular clusters. These tables shall be made available electronically.

Furthermore, we investigate the Oosterhoff dichotomy \citep{Oosterhoff1939,Oosterhoff1944} that divides the Milky Way (MW) globular clusters based on their RR~Lyrae stars content. The Oosterhoff type I (Oo\,I) clusters have higher metallicities ([Fe/H]$> -1.5$ dex) and their RR~Lyrae variables have on average shorter pulsation periods ($<P_{\rm RRab}> \approx 0.55$\,day). On the other hand, Oosterhoff type II (Oo\,II) globular clusters are more metal-poor ([Fe/H]$< -1.5$ dex) containing RR~Lyrae pulsators with on average longer pulsation periods ($<P_{\rm RRab}> \approx 0.65$\,day).

In recent decades, various theories have been suggested to explain the Oosterhoff dichotomy. One such theory is the presence of a hysteresis zone \citep{vanAlbada1973,Caputo1978}, which indicates that RR Lyrae stars pass through the instability strip (IS) twice. They proposed that Oo\,II RR~Lyrae stars transition from the blue edge to the red edge of the IS, whereas Oo\,I RR~Lyrae stars move in the reverse direction, from red to blue. Another proposition put forward suggested that a higher helium content in Oo\,II clusters could account for the differences between Oosterhoff groups \citep{Sandage1981a,Sandage1981b}. 
Recent studies of field RR~Lyrae stars showed a smooth transition in periods as a function of spectroscopic metallicity \citep{Fabrizio2019,Fabrizio2021}. This suggests that the observed Oosterhoff type in MW globular clusters is a result of the deficiency of metallicity-intermediate clusters with RR~Lyrae stars. Lastly, the origin of the Oosterhoff dichotomy observed in Galactic bulge RR Lyrae stars (e.g., Prudil et al. 2019a,b) remains unclear. Specifically, the two Oosterhoff groups form distinct hook-like structures in Fourier parameter space (in the ratio of Fourier amplitudes and phase differences $R_{\rm 31}$ vs. $\varphi_{21}$ and $\varphi_{31}$). These structures are observed only in non-modulated fundamental-mode RR~Lyrae stars in the bulge \citep[stars without the Blazhko effect;][]{Blazhko1907} and can be distinguished in the period-amplitude plane, similar to the Oosterhoff groups. 

Dedicated photometric studies of individual globular clusters have revealed that 
the distribution of the RR Lyrae pulsating modes carries some characteristics that distinguish
Oo I from Oo II groups \citep{Yepez2022}. The instability strip is in fact the union of the fundamental mode and the first overtone instability strips \citep{Bono1994}. The intersection of these two strips is known as the inter-order or "either-or" region. By compiling the distributions of RRab and RRc stars in CMDs from previous accurate photometric studies, \citet{Yepez2022} showed that in Oo II type clusters the fundamental mode RRab stars do not occupy the inter-order region, that is, they sit to the red of the first overtone red edge (FORE) of the strip, whereas this situation may or may not occur in Oo I clusters, where the inter-order region may or may not be shared by the RRab and the RRc stars. While this can be reconciled with the hysteresis paradigm described above, it is also of relevance when considering the Oosterhoff gap on the plane defined by the mean cluster metallicity and the HB structural parameter. In the present work, we shall address these perspectives of the Oosterhoff dichotomy.

\section{The strategy} \label{sec:strategy}

The CVSGC catalog provides equatorial coordinates of nearly all listed variable stars. For a small subset of variables only pixel coordinates were available and these objects were not used in this analysis. The remaining stars with equatorial coordinates and equinox were further used in membership assessment. To that end, we used the catalog of globular cluster stars assembled by \citet{Vasiliev2021}\footnote{Available in the electronic form here \url{https://zenodo.org/records/4891252}.} using astrometric solutions obtained from the \textit{Gaia} early data release 3 \citep[EDR3,][]{Lindegren2021} catalog. This catalog offers membership probabilities, $p_{\rm mem}$, to individually analyzed stars within a given radius around the cluster. Using the proper motions of individual stars in this catalog we transformed their coordinates from \textit{Gaia} equinox (J2016) to an appropriate equinox matching the one listed in the CVSGC catalog (J2000 or J1950). The modified membership catalog was then crossmatched with individual cluster variables (using a $2$\,arcsec radius), and for the best matching counterpart in the CVSGC, the membership probability was acquired. 

\begin{figure}
\includegraphics[width=\columnwidth]{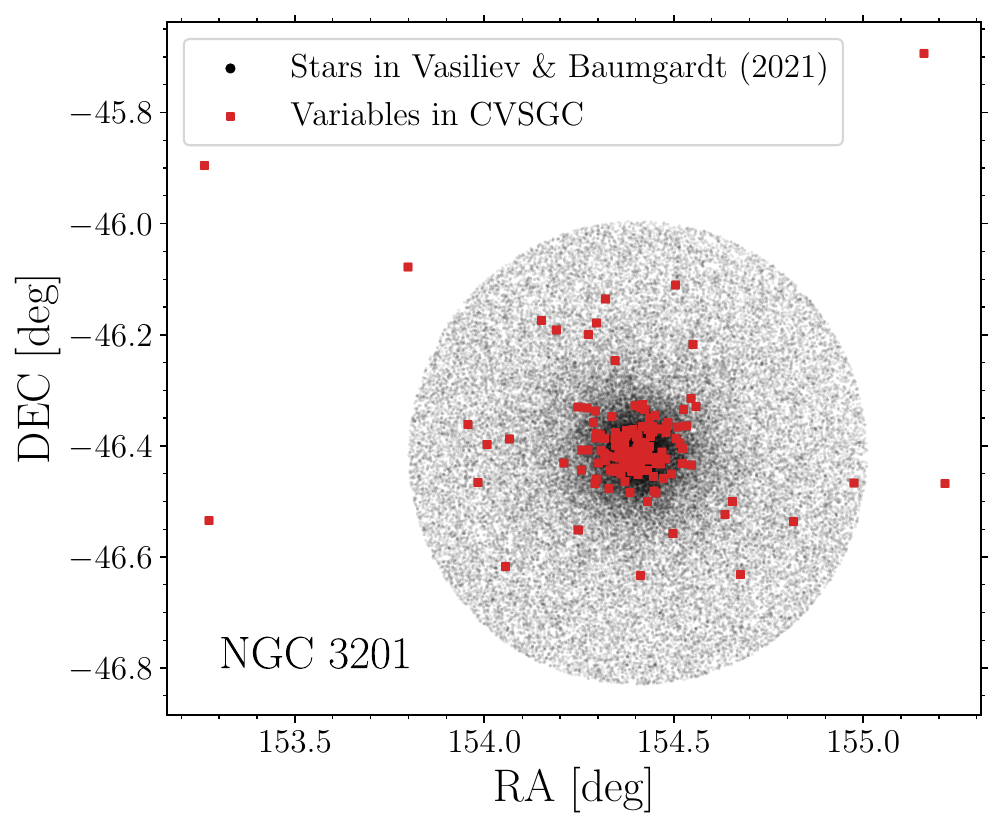}
\caption{The distribution of equatorial coordinates of stars in the vicinity of the NGC-3201. The black points represent objects with membership probabilities from \citet{Vasiliev2021}, and the red squares stand for variables from the original CVSGC catalog.}
\label{fig:NGC3201}
\end{figure}

In some cases, where we could not find a match of a given variable in the membership catalog we marked such an object with $p_{\rm mem}$ equal to $-999$. The same applies to cases where a variable object was located outside the radius analyzed by \citet{Vasiliev2021} see Figure~\ref{fig:NGC3201} for an example of such a case. For the updated version of the CVSGC catalog we chose the following nomenclature scheme; variables with $p_{\rm mem} \geq 0.85$ marked with \texttt{m1}, for $0.7 \leq p_{\rm mem} < 0.85$ we used \texttt{m2}, stars with    $0.5 \leq p_{\rm mem} < 0.7$  we assigned \texttt{m3} and lastly stars with $p_{\rm mem} < 0.5$ we marked as \texttt{f}. As mentioned above, for stars that did not have a counterpart in \citet{Vasiliev2021} catalog, were outside the analyzed radius, or did not have equatorial coordinates in CVSGC we used label \texttt{U}.

Here we will highlight some of the cases where including membership had a large impact on the amount of RR~Lyrae stars in a given cluster. Specifically, clusters located in the Galactic bulge NGC~6441, NGC~6266 and NGC~6715. For NGC~6441 the reason is the large number of non-members in the line-of-sight toward this cluster associated with the Galactic bulge. The NGC~6266 on the other hand has many RRab stars not included in the \citet{Vasiliev2021} catalog (only $83$ are included from the total of $141$ RRab listed in CVSGC). In the case of NGC~6715, a combination of both foreground RR~Lyrae stars and variables not included in the catalog by \citet[][either due to the conditions on \textit{Gaia} astrometry or some RR~Lyrae stars were not located in the radius for membership search]{Vasiliev2021}.

\section{RR~Lyrae stars in Globular clusters} \label{sec:RRLinClus}

The catalog of variable stars enhanced with membership probabilities permits us to explore some of the properties of the globular clusters connected to their variable star content. In this study, we concentrate on the Oosterhoff dichotomy \citep{Oosterhoff1939,Oosterhoff1944}. This almost a-century-old phenomenon, present in the MW globular clusters containing RR~Lyrae variables, has been thoroughly explored in the work by \citet{Catelan2009}. Here we aim to provide an update on the current standing of the MW globular clusters and their RR~Lyrae content. 

We will particularly focus on Figure~5 presented in a seminal work by \citet{Catelan2009}, and on the so-called Oosterhoff gap. This Figure depicts the average pulsation periods of the fundamental-mode RR~Lyrae stars associated with a given globular cluster vs. the cluster's metallicity. This plane is the cornerstone for identifying and examining the Oosterhoff dichotomy among MW globular clusters. In our case, the estimation of average pulsation periods of RRab stars linked with a given globular cluster was trivial. Obtaining the metallicity information, on the other hand, is complex. The CVSGC catalog contains pulsation periods of a given RR~Lyrae star but it does not have any information on their metallicity nor on their Fourier coefficients \citep[often used to estimate photometric metallicities of RR~Lyrae stars, see, e.g.,][]{Jurcsik1996,Dekany2021,Arellano2024}. To supplement our RR~Lyrae globular cluster dataset with metallicities we used two sources, the first source was the set of [Fe/H]$_{\rm AF}$ values obtained via the RRab and RRc light curve Fourier decomposition reported by \citet{Arellano2024} (hereinafter AF24) for a sample of $39$ globular clusters. These photometrically obtained metallicities are given in the metallicity scale of \citet{Zinn1984} and were subsequently transformed into the spectroscopic scale of \citet{Carretta2009}. These values we have named as [Fe/H]$_{\rm AF}$ in Fig. \ref{fig:FePab}. The second source of metallicities is based on photometric information (pulsation periods and Fourier phase difference $\varphi_{31}$) from the \textit{Gaia} DR3 RR~Lyrae catalog \citep{Clementini2023}. We crossmatched the \textit{Gaia} RR~Lyrae catalog with cluster members from \citet{Vasiliev2021} using their \texttt{source\_id}'s. We assumed only RR~Lyrae with $p_{\rm mem} > 0.5$ as members and for those, we estimated photometric metallicities based on relations from \citet[][from hereon noted as K24]{Kunder2024}. For the given cluster we then estimated the average photometric metallicity and used it in our comparison. We note that these metallicities are on the metallicity scale defined by \citet{Crestani2021,For2011,Chadid2017,Sneden2017}, abbreviated CFCS. Here is important to emphasize that we used \citet{Clementini2023} catalog only to estimate the average photometric metallicities for a given cluster but not the average pulsation periods. For average pulsation periods, we used the CVSGC after removing RRab stars that were likely not members \citep[based on catalog of][]{Vasiliev2021}.

We note that for globular clusters associated with the Oosterhoff group III (NGC~6441 and NGC~6388), the photometric metallicities do not accurately reflect the spectroscopic metallicities. For example, in NGC~6441 spectroscopic metallicities yield [Fe/H]$=-0.41 \pm 0.06$\,dex \citep{Clementini2005} while photometric metallicities give [Fe/H]$_{\rm phot}=-1.95 \pm 0.47$\,dex or [Fe/H]$_{\rm AF}=-1.23\pm 0.17$\,dex. The photometric metallicities of NGC~6388 suffer a similar discrepancy [Fe/H]$=-0.48 \pm 0.08$\,dex \citep[based on red giants][]{Carretta2022} and [Fe/H]$_{\rm phot}=-1.91 \pm 0.14$\,dex, or [Fe/H]$_{\rm AF}=-1.23\pm 0.06$\,dex using our RR~Lyrae stars.

\subsection{The Oosterhoff gap and the extragalactic merging events.} \label{subsec:OoGap}

\citet{Oosterhoff1939,Oosterhoff1944} noticed that the average of periods of the RR~Lyrae variables of the RRab-type located in the MW globular clusters was either $\sim 0.55$ days (the Oo I clusters) or $\sim 0.65$ days (the Oo II clusters). In his review \citet{Catelan2009}, discussed this Oosterhoff dichotomy in the Galactic halo cluster where a sharp separation between Oo I and Oo II is clear and noticed that in very few clusters the RRab stars have periods in the range $ 0.58 - 0.62$ days, this range is called the Oosterhoff gap. Furthermore, \citeauthor{Catelan2009} remarks that dwarf globular spheroidal galaxies (dSph), their clusters or other clusters of clear extragalactic nature, like those in the LMC, preferential occupy the Oosterhoff gap.  The Oosterhoff gap is also prominent in the plane $HBt$-[Fe/H], being $HBt$ the HB structure parameter defined as $HBt=(B-R)/(B+V+R)$ \citep{Lee1994} where $B$ and $R$ are the number o stars to the blue and red of the instability strip (IS) respectively, while $V$ is the number of RR~Lyrae stars in the IS.
On this plane, \citet{Catelan2009} identified a "forbidden" region systematically devoided of Galactic clusters but again, preferentially occupied by systems of extragalactic nature.

These results are of relevance in the identification of GCs that were born $in-situ $ in the MW and those that were accreted later during merging events of smaller stellar systems, like dwarf galaxies and globular clusters. The identification of extragalactic mergers with the MW and studying their impact on the stellar distributions, such as stellar halo substructures is an interesting and challenging field and constitutes the discussion of many recent investigations. Comparison of observed GCs distribution with hydrodynamical simulations of GCs formation and evolution, splitting of the energy angular momentum space, and age-energy considerations, have allowed several authors to associate families of GCs with identified accretion events. We shall refer to the work by \citet{Callingham2022} and will follow their mergers identifications and nomenclature as well as the association of GCs to each of these mergers via a thorough chemo-dynamical analysis. The interested reader in the major contributions to the field in recent years can benefit from the detailed introduction to the paper by \citet{Callingham2022}.

The merger events and the probable association of GCs to them are summarized in Table 1 of \citet{Callingham2022}. The identified groups are the $in-situ$ formation Galactic bulge and disc; and the mergers \textit{Gaia}-Enceladus-Sausage (GES), Helmi, Kraken, Sagittarius, and Sequoia. We shall follow this table in the construction of the 
$<P_{\rm ab}>$-[Fe/H] and $HBt$-[Fe/H] planes to be discussed in the following sections.

\begin{figure*}
\begin{center}
\includegraphics[width=14.5cm]{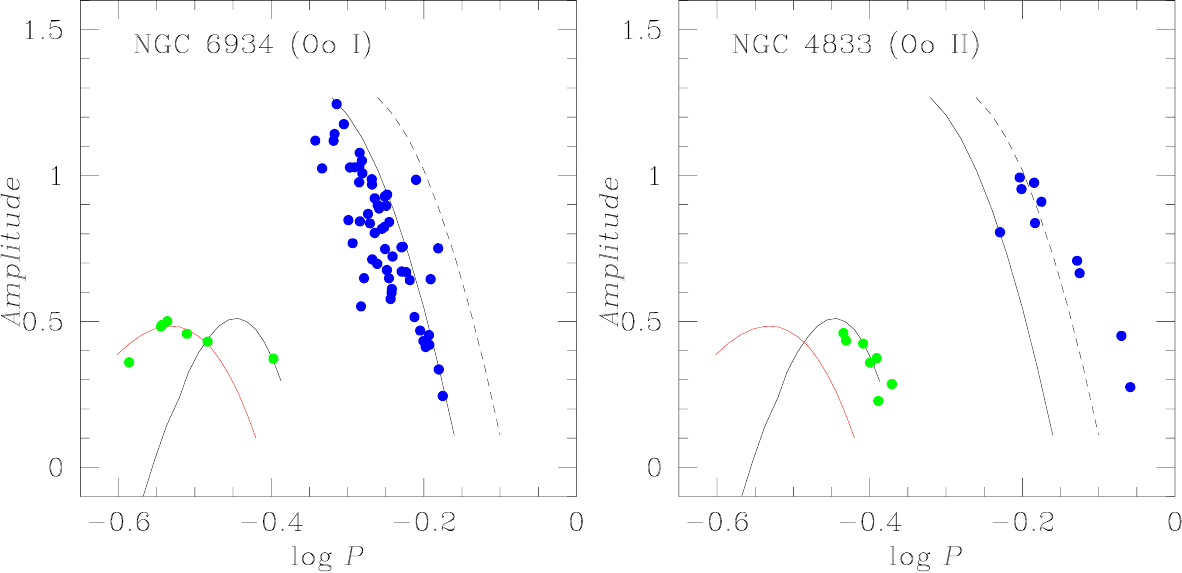}
\caption{Two examples of the log~$P$~-~Amplitude plane for clearly defined Oo~I and Oo~II clusters. Blue and green circles represent RRab and RRc stars respectively. The loci for unevolved and evolved RR Lyraes were defined by \citet{Cacciari2005} and by \citet{Kunder2013a} for the RRab and by \citet{Kunder2013b} and by \citet{Yepez2020} for the RRc stars.}
\label{fig:Bailey}
\end{center}
\end{figure*}

Categorizing individual clusters into the Oosterhoff groups was done in two steps, based on visual inspection of the period-amplitude diagram for each cluster using only stars associated with the given cluster \citep[while using pulsation periods and amplitudes from the \textit{Gaia} RR~Lyrae catalog][]{Clementini2023}. In the log $P$---amplitude diagram there are known clear sequences for unevolved and evolved RRab \citep{Cacciari2005,Kunder2013a} and RRc \citep{Kunder2013b,Yepez2020} stars typical of Oo~I and Oo~II clusters. Hence, the cluster member RR~Lyrae distribution on this plane is generally a good indicator of the Oosterhoff type.  In Fig.~\ref{fig:Bailey} we offer examples of the log~$P$~-~amplitude plane in two clusters with their Oosterhoff type clearly defined. This can be complemented by the fact that Oo~I clusters have a period average of their RRab, $<P_{ab}> \leq 0.60$ d and [Fe/H] $\gtrsim -1.5$ stars. In most clusters, the above criteria are clear and sufficient to assign an Oosterhoff type to a given cluster. However, a few clusters remain dubious from the log $P$---amplitude diagram criterion, generally due to the low number of RR Lyrae stars involved. We opted for leaving their Oo type entry with a question mark in Table \ref{FevsPab} and they are distinguished by plotting them with open triangles in Figs. \ref{fig:FePab} and \ref{fig:FeL}.

\subsection{The $<P_{\rm ab}>$-[Fe/H] plane}
\label{PabFeH}

In three panels of Figure~\ref{fig:FePab} we displayed our results for three distinct metallicity sources. In panel $b$, we see a clear separation between Oosterhoff groups, and with red lines we mark the approximate position of the Oosterhoff gap. We note that globular clusters NGC~6402 and NGC~6522 are located in the Oosterhoff gap albeit with an uncertain average pulsation periods. Panel $a$ offers a slightly different view. Due to the larger sample size of globular clusters with RRab members, we seem to populate the Oosterhoff gap previously devoid of clusters in this diagram. For completenes, in panel $c$ the same plane is displayed employing the spectrocopic UVES iron abundances from the work of \citet{Carretta2009}, listed in their table A.1.

There are several reasons why we do not see the Oosterhoff gap in panels $a$ and $c$ of Fig.~\ref{fig:FePab}. First, our dataset of MW globular clusters ($79$ clusters) is larger than that one used in \citet[][$41$ clusters]{Catelan2009}. Thus, potentially additional clusters populate the Oosterhoff gap. This is the case for NGC~5946, NGC~6235, NGC~6522, NGC~6638, and Ter~1 which were not included in \citet{Catelan2009}. These clusters in our analysis have eight or fewer RRab variables. The other three globular clusters in the Oosterhoff gap, NGC~6715 (M54), NGC~6402 (M14), and NGC~6864 (M75), are much more populated by fundamental-mode pulsators ($96$, $42$, and $20$, respectively). The NGC~6402 and NGC~6864 clusters differ in our analysis from the analysis by \citet{Catelan2009} partially in the average pulsation period, NGC~6402, $<P_{\rm ab}^{\rm This~study}> = 0.599$\,day and $<P_{\rm ab}^{\rm Catelan+2009}> = 0.564$\,day, and partly in metallicity, NGC~6864, [Fe/H]$_{\rm phot}^{\rm This~study}=-1.385$\,dex and [Fe/H]$_{\rm phot}^{\rm Catelan+2009}=-1.16 $\,dex. Lastly, globular clusters previously associated with MW dwarf galaxies, NGC~6715, NGC~4147, NGC~5634, NGC~1851, NGC~2808, NGC~5286 are also included in the Fig.~\ref{fig:FePab}, but only NGC~6715 (M54) falls into the Oosterhoff gap.

\subsection{The HB morphology vs [Fe/H]}
\label{HByFeH}

The classical $HBt$ parameter is convenient to describe the morphology of the HB since it only involves star counts and it is easy to calculate. However it also carries some inconveniences commented by \citet{Torelli2019}, such as the saturation for metal-rich cluster ([Fe/H] > -1.0 ) at $HBt \sim$ -1, as well as for metal-poor ([Fe/H] < -1.8 ) clusters at at $HBt \sim $ +1. We shall explore the nature of Oosterhoff gap in two morphology structure vs [Fe/H] planes. First the $HBt$-[Fe/H] plane displayed in Fig. \ref{fig:FeL}. To build this plane we proceeded as follows. The metallicity values were taken from the work of AF24. These values were calculated via the Fourier decomposition of light curves and ad-hoc calibrations and zero points for RRab and RRc stars, hence the metallicity values, duly transformed to the spectroscopic scale [Fe/H]$_{\rm UVES}$ \citep{Carretta2009}, are independent and were averaged. The good agreement between these Fourier estimates and the spectroscopic metallicities of \citet{Carretta2009} has been demonstrated in the paper of AF24, hence for clusters not included in the list of AF24, their values were taken from \citet[][their table A.1]{Carretta2009}. These metallicities are listed column 6 of Table~\ref{FevsPab} as [Fe/H]$^{\rm UVES}_{\rm AF}$.

Similarly, the HB structure parameter $HBt$ values were collected from AF24 where their calculation involved cluster membership considerations. For clusters not included in the work of AF24, $HBt$ was taken from the work of \citet[][their table 1]{Torelli2019} or otherwise from \citet[][Tables 2 and 3]{Catelan2009}. For the sake of clarity all the used values are listed in column 14 of Table~\ref{FevsPab}.

\citet{Torelli2019} amply discussed other HB morphology indexes, their pros and cons and introduce the $\tau_{HB}$ index which is calculated as the ration of the areas subtended by the cumulative number distributions along the HB in magnitude $I$ and colour $(V-I)$. $\tau_{HB}$ has a larger dynamical range than $HBt$ and does not saturate. We find of interest to note the Oosterhoff gap morphology in the $\tau_{HB}$-[Fe/H] plane, which is displayed in Fig. \ref{tauFeH}. To build this plane we adopted the $\tau_{HB}$ values from table 2 of \citet{Torelli2019} for 46 clusters in common with those in our Table \ref{FevsPab}, while the metallicities, symbols and colour code are those as in Fig. \ref{fig:FeL}. The two segmented lines have been placed by eye and may be a bit arbitrary however enhance the clear presence of the Oosterhoff gap. The position of some clusters inside the gap is very marginal at most. One may compare in Figs. \ref{fig:FeL} and \ref{tauFeH} the positions of NGC 4147, NGC 4590, NGC 5053, NGC 6426 and NGC 5286. Unfortunatelly there are no estimations of $\tau_{HB}$ for NGC 6402, NGC 6558 and NGC 6715, apparently inside the gap in Fig. \ref{fig:FeL}.

\section{Discussion}
\label{sec:DISCUSSION}

Looking at our analysis and study by \citet{Catelan2009} as a whole, the determined average pulsation periods differ only mildly (dispersion of the difference equal to $0.03$\,day). The two largest differences occur for NGC~6558 and NGC~6642. In a recent detailed study on NGC~6558 \citep{Arellano2024b}, five RRab members were identified, although only three were recovered here due to the low membership probability assigned to V1 and V4 by \citet{Vasiliev2021}. Depending upon the inclusion of these two variables the value of <P$_{ab}$> is either $0.60$ or $0.67$\,days. For NGC~6642 we found only two RRab members while in \citet{Catelan2009} were reported ten. In both cases, the large differences are likely caused by the low number of RRab stars.


The following systems deserve a comment; IC~4499, NGC~6284, NGC~6864, Rup~106 and NGC~6558 on their Oosterhoff classification in our study. The first four globular clusters were classified as Oosterhoff intermediate in \citet{Catelan2009}, based on our classification approach (described above) we classified them as Oosterhoff type I clusters. The NGC~6558 was not classified into Oosterhoff groups due to a low number of RRab stars (only 3-5 stars associated) \citep[see][]{Arellano2024b}.

A note is in order regarding the $HBt$ value of NGC~6626. From the log $P$---amplitude diagram, it is clear that we are dealing with an Oo I cluster, however, the HB structural parameter $HBt$=+0.90  reported by \citet{Catelan2009} would put NGC~6626 among the Oo II clusters. We have used the \textit{Gaia} CMD for cluster members and calculated $HBt$=+0.43. We have used this value in this work.

\begin{figure*}
\begin{center}
\includegraphics[width=2\columnwidth]{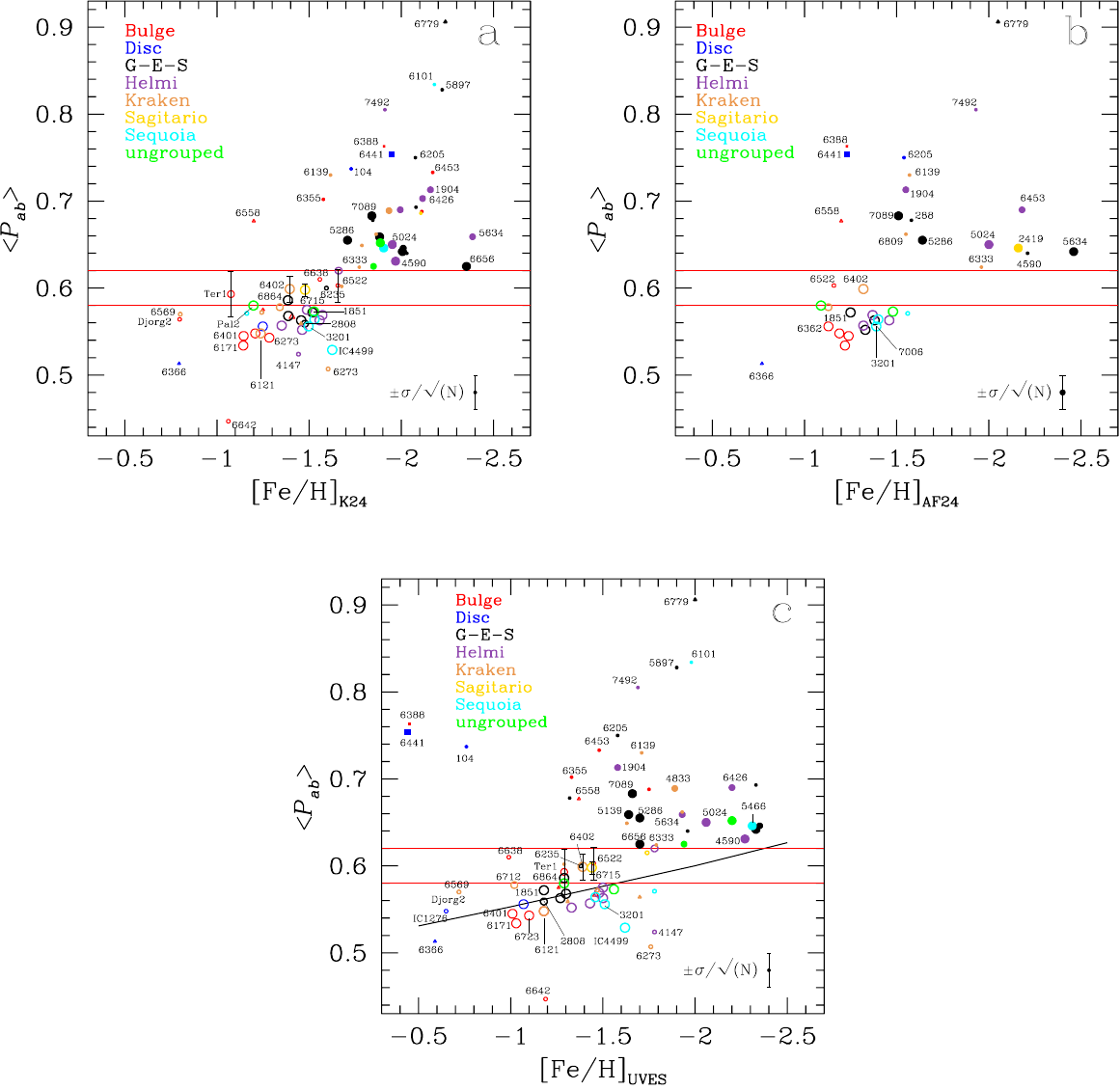}
\caption{The mean photometric metallicities versus the average period of cluster member RRab stars in a sample of Galactic globular clusters. Open and filled circles represent Oo I and Oo II clusters respectively. Filled squares are employed for Oo III clusters and open triangles for clusters with unclear Oosterhoff type from their log $P$-amplitude diagram. In Panel a) a sample of $79$ systems is considered with the values of [Fe/H] based on the calibration of \citet{Kunder2024}. In Panel b) the iron abundances for a sample of $39$ globulars are from the Fourier light curve decomposition reported by \citet{Arellano2024}. In Panel c) the UVES iron abundances for a sample of 78  clusters are those from \citet[][thier table A.1]{Carretta2009}. The black solid line represents the $<P_{\rm ab}>$ vs. metallicity trend based on field RR~Lyrae stars from \citet[][thier equation 7]{Fabrizio2021}. Clusters associated with different dwarf galaxy mergers to the Milky Way are plotted with different colours. The associations are those defined by \citet[][their Table 1]{Callingham2022}. Three sizes are used according to the number of member RRab stars considered in each cluster; for N$\leq 5$, $5 < N \leq 10$, and $10 < N$. The selected error bars are estimates of the confidence interval calculated as $\sigma/ \sqrt(N)$. Some clusters are not labeled due to crowdedness; their identification and coordinates can be found in Table~\ref{FevsPab}. The two horizontal red lines suggest the limits of the Oosterhoff gap according to \citet{Catelan2009}. See Subsection~\ref{PabFeH} for a detailed discussion of these figures.}
\label{fig:FePab}
\end{center}
\end{figure*}

\begin{figure*}
\begin{center}
\includegraphics[width=16cm]{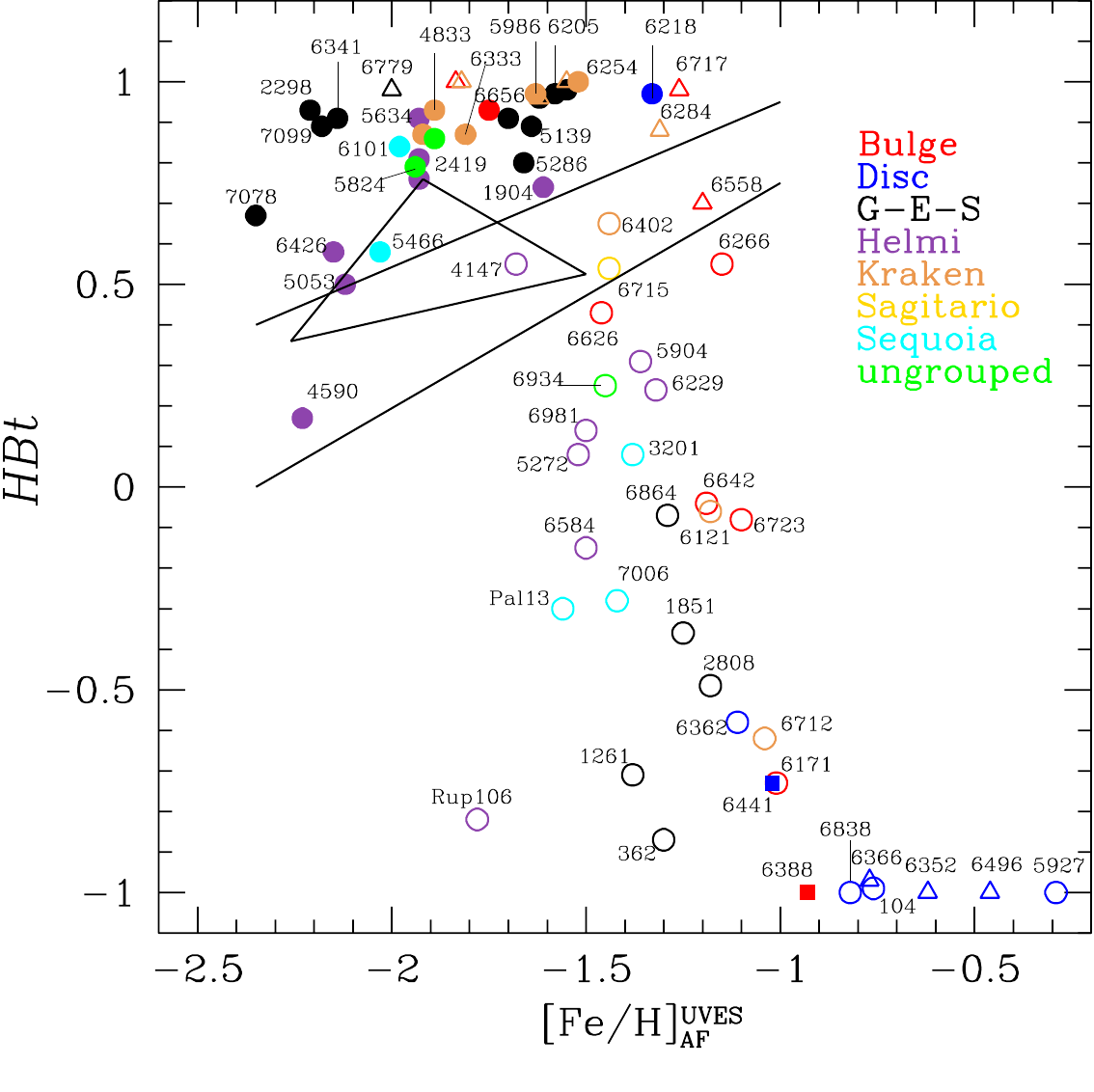}
\caption{The metallicities in the UVES scale \citep{Carretta2009} versus the HB type parameter $HBt$. The data sources are described in \S \ref{HByFeH}. Symbols are as coded in the caption of Fig. \ref{fig:FePab}. The two straight lines represent the theoretical limits of the Oostherhof gap according to \citet{Bono1994}. The triangle marks the identified region devoided of Galactic clusters according to \citet{Catelan2009} (see $\S$ \ref{subsec:OoGap} and \ref {sec:DISCUSSION} for a discussion). A few clusters remained unlabeled due to crowdedness; their identification and coordinates can be found in Table \ref{FevsPab}.}
\label{fig:FeL}
\end{center}
\end{figure*}

\begin{figure*}
\begin{center}
\includegraphics[width=16cm]{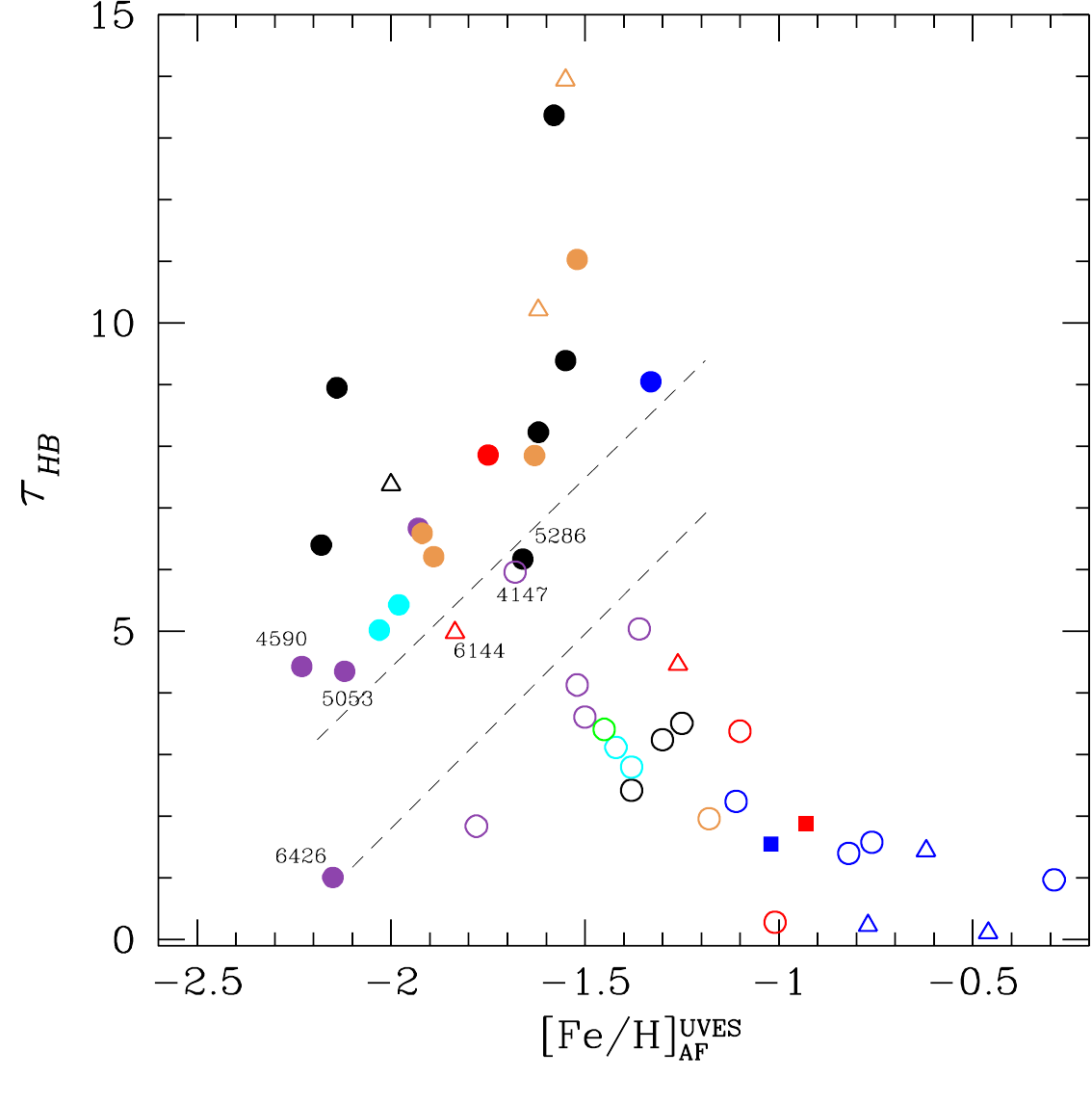}
\caption{The metallicities in the UVES scale \citep{Carretta2009} versus the HB morphology parameter $\tau_{HB}$ \citep{Torelli2019}. Symbols and colour code is as in Fig. \ref{fig:FeL}. The segmented lines mark the approximate limits of the Oosterhoff gap.}
\label{tauFeH}
\end{center}
\end{figure*}

In his analysis of the $<P_{\rm ab}>$-[Fe/H] and $HBt$-[Fe/H] planes, \citet{Catelan2009} noticed that the Oosterhoff gap and the so-called "forbidden" region or "Oosterhoof gap?" (marked with a triangular area in Fig.~\ref{fig:FeL}), were devoided of Galactic globular clusters but was otherwise populated by dSph galaxies and their associated clusters orbiting the MW. In present Figs. \ref{fig:FePab},  \ref{fig:FeL}  and \ref{tauFeH} we analyze the distribution of globular clusters presently in the MW, potentially formed $in-situ$ (Bulge and Disc clusters) and those that are associated with extragalactic mergers \citep{Callingham2022}. There are two striking pieces of evidence in these plots; a) the Oosterhoff gap in the plane $<P_{\rm ab}>$-[Fe/H] is populated by numerous clusters associated with extragalactic mergers to the MW (see $\S$~\ref{PabFeH}), b) the vast majority of GC's associated to extragalactic mergers are clearly off the Oosterhoff gap, as well as off they are from the "forbidden region" in the $HBt$-[Fe/H] plane. This result is confirmed with the use of the $\tau_{HB}$ morphology index in the plane $\tau_{HB}$-[Fe/H]. We should note that NGC~6402 (M14) and NGC~6715 (M54), associated with the Kraken/Heracles \citep{Kruijssen2020,Horta2021} and Sagittarius mergers respectively, are the only clusters that by their average period $<P_{\rm ab}>$ and their HB structure $HBt$ do fall in the Oostherhoff gap.

The case of M14 is particularly interesting. It has been shown by \citet{Massari2019} and \citep{Callingham2022} that NGC~6402 low orbital-energy space linked with the Kraken/Heracles merger event. While the $<P_{\rm ab}>$-[Fe/H] and $HBt$-[Fe/H] correlations may not be enough to distinguish an extragalactic from an $in-situ$ cluster, at least in the case of M14 this conclusion is consistent with the chemo-dynamical analysis of \citep{Callingham2022}.

A recent study by \citet{Luongo2024} examined the Oosterhoff dichotomy from a kinematic perspective, focusing on field RR Lyrae stars and globular clusters using integrals of motion. Their analysis showed that field in-situ RR~Lyrae stars differ from those associated with the accreted population in the period-amplitude plane, suggesting that the Oosterhoff dichotomy among field RR Lyrae stars was introduced by past accretion events. Our analysis of the Oosterhoff dichotomy, using the updated RR Lyrae catalog for Milky Way globular clusters, based on the $<P_{\rm ab}>$-[Fe/H] and $HBt$-[Fe/H] planes, offers a slightly different perspective.

For example, the globular clusters NGC~6715 and NGC~6402, currently associated with ongoing and past mergers, fall into the Oosterhoff gap, which is predominantly populated by globular clusters from dwarf galaxies. This suggests an import of globular clusters associated with the intermediate Oosterhoff category, rather than Oosterhoff I or II. One potential explanation for the differences between our conclusions and those of \citet{Luongo2024} is the distinction between the dwarf galaxies currently observed in the Milky Way and those that were accreted in the past.

In particular, the past mergers associated with GES, Helmi streams and Kraken/Heracles appear to have stellar masses around $10^{8}$\,M$_{\odot}$  and higher \citep{Horta2021,Koppelman2019,Lane2023}, while current dwarf galaxies have masses around $10^{7}$\,M$_{\odot}$ or lower \citep[see][their Table~1]{Kirby2013}. The more massive accreted dwarfs could have experienced prolonged and more efficient star formation, leading to higher metallicities, which may have pushed some of their globular clusters into the Oosterhoff I group. This could led to the depopulation of the Oosterhoff gap and the emergence of the Oosterhoff dichotomy among the field-accreted RR Lyrae stars.


\begin{table*}
\caption{Metallicities and period averages of cluster member RR~Lyrae stars in a sample globular clusters. The association of each cluster to dwarf galaxy mergers to the Milky Way is indicated in the last column \citep{Callingham2022}.}
\begin{center}
\label{FevsPab}
	\begin{tabular}{lccccccccrccrcc}
			\hline
  Name     & Oo Type &[Fe/H]$_{\rm K24}$& s& N & [Fe/H]$^{\rm UVES}_{\rm AF}$ & s & $<P_{\rm ab}>$ & s &  N & $<P_{\rm c}>$ &s & N&$HBt$ &Group$^*$\\    
            \hline
  Arp2     & I &    ---    &  ---  & 0   &  ---   & ---  & 0.615  & 0.115 & 4   &  0.291 &---  & 1&---&Sg\\   
  BH140    & II &    -1.716 &  0.023& 2   &  ---   & ---  & ---    & ---   & 0   &  ---   &---  & 0&---&GES\\    
  BH176    & I &   ---     & ---  & 0    & ---   &---   & ---    & ---   &0    & ---    &---  & 0&---&Di\\   
  BH261    & ? &  ---    &  ---  & 0   &  ---   & ---  & ---    & ---   & 0   &  ---   &---  & 0&---&Kr\\    
  Djorg 2   & I &    -0.798 &  0.398& 1   &  ---   & ---  & 0.564  & ---   & 1   &  0.303 &---  & 1&---&Bl\\    
  ESO 452   & I &    ---    &  ---  & 0   &  ---   & ---  & ---    & ---   & 0   &  ---   &---  & 0&---&Bl\\    
  FSR1716  & I &   -1.328  & 0.174  &5    & ---    &---   &---     &---   &0    & ---    &--   &0 &---&Kr\\   
  FSR1735  & II &    ---    &  ---  & 0   &  ---   & ---  & ---    & ---   & 0   &  ---   &---  & 0&---&Kr\\    
  FSR1758  & ? &    -1.933 &  0.291& 3   &  ---   & ---  & ---    & ---   & 0   &  ---   &---  & 0&---&Kr\\    
  HP1      & I &    -1.226 &  0.100& 5   &  ---   & ---  & ---    & ---   & 0   &  ---   &---  & 0&&Bl\\    
  IC1276   & I &    ---    &  ---  & 0   &  ---   & ---  & 0.548  & ---   & 1   &  ---   &---  & 0&---&Di\\    
  IC4499     &I  &   -1.625  & 0.217 &57   & ---    &---   &0.529   &0.108 &75   & 0.356  &0.006 &11&---&Sq\\   
  Laevens3 & I &    -1.751 &  0.141& 2   &  ---   & ---  & ---    & ---  & 0   &  ---   &---  & 0&---&Ug\\    
  Lynga7   & I &    ---    &  ---  & 0   &  ---   & ---  & ---    & ---  & 0   &  ---   &---  & 0&---&Di\\    
  NGC104    & II &    -1.728 &  ---  & 1   &  ---   & ---  & 0.737  & ---  & 1   &  ---   &---  & 0&---&Di\\    
  NGC288    & II &    -1.845 &  ---  & 1   &  -1.55 & ---  & 0.678  & ---  & 1   &  0.403 &---  & 1&+0.98&GES\\    
  NGC5286    &II  &   -1.708  & 0.092 &3    & -1.66  &0.15  &0.655   &0.095 &16   & 0.344  &0.038 &11&+0.80&GES\\  
  NGC362    & I &    -1.387 &  0.224& 14  &  -1.30   & ---  & 0.568  & 0.060& 23  &  0.365 &0.039& 3&-0.87&Di\\    
  NGC1261   & I &    -1.457 &  0.127& 9   &  -1.38 & 0.05 & 0.563  & 0.047& 14  &  0.315 &0.020& 6&-0.71&GES\\    
  NGC1851    &I  &   -1.519  & 0.307 &16   & -1.25  &0.15  &0.572   &0.063 &21   & 0.307  &.04  &11&-0.36&GES\\  
  NGC1904   &II &    -2.158 &  0.172& 2   &  -1.61 & 0.14 & 0.713  & 0.060& 6   &  0.329 &0.007& 3&+0.74&He\\   
  NGC2298   & II &    -2.029 &  ---  & 1   &  -2.21 & 0.09 & 0.640  & ---  & 1   &  0.369 &0.002& 2&+0.93&GES\\   
  NGC2419    &II  &   -1.886  & 0.211 &10   & -1.89    &0.28  &0.652   &0.067 &32   & 0.371  &.035 &32&+0.86&Ug\\   
  NGC2808   & I &    -1.477 &  0.094& 3   &  -1.18  & ---  & 0.559  & 0.032& 6   &  0.303 &0.35 & 5&-0.49&GES\\   
  NGC3201   & I &    -1.497 &  0.137& 70  &  -1.38 & 0.10 & 0.556  & 0.045& 75  &  0.327 &0.035& 7&+0.08&Sq\\    
  NGC4147    &I  &   -1.442  & 0.050 &5    & -1.68    &0.26  &0.524   &0.043 &5    & 0.321  &.044 &16&+0.55&He\\   
  NGC4372   & I &    ---    &  ---  & 0   &  ---   & ---  & ---    & ---  & 0   &  ---   &---  & 0&---&Kr\\   
  NGC4590    &II &   -1.969  & 0.187 &10   & -2.23    &0.06   &0.631   &0.063 &13   & 0.367  &.025 &15&+0.17&He\\  
  NGC4833   & II & -1.933 &  0.198& 10  &  -1.89   & 0.05  & 0.689  & 0.082& 9   &  0.384 &0.036& 8&+0.93&Kr\\   
  NGC5024    &II  &   -1.951  & 0.183 &25   & -1.93  &0.06  &0.650   &0.065 &27   & 0.035  &.038 &32&+0.81&He\\   
  NGC5053   & II &    -1.994 &  0.162& 5   &  -2.12 & 0.14 & 0.690  & 0.061& 6   &  0.346 &0.033& 4&+0.50&He\\    
  NGC5139 $\omega$Cen &II  &   -1.882  & 0.362 &22   &-1.64   &0.09   &0.659   &0.116 &70   & 0.360  &.051 &82&+0.89&GES\\   
  NGC5272 M3  &I  &   -1.557  & 0.207 &143  & -1.52  &0.16  &0.563   &0.062 &164  & 0.325  &.040 &41&+0.08&He\\   
  NGC5286    &II  &   -1.708  & 0.092 &3    & -1.66  &0.15  &0.655   &0.095 &16   & 0.344  &.038 &11&+0.80&GES\\   
  NGC5466   & II &    -1.904 &  0.217 &11  &  -2.03 & 0.14 & 0.646  & 0.080 & 13  &  0.358 &0.045& 7&+0.58&Sq\\    
  NGC5634   & II &    -2.386 &  0.567 &1   &  -1.93   & 0.09  & 0.659  & 0.078 & 9   &  0.341 &0.043& 6&+0.91&He\\    
  NGC5824   & II &    -1.849 &  0.228 &10  &  -1.94   & 0.14  & 0.625  & 0.022 & 6   &  -1.94   &0.14  & 0&+0.79&Ug\\    
  NGC5897   & II &    -2.221 &  0.125 &3   &  -1.90   & 0.06  & 0.828  & 0.024 & 3   &  0.445 &0.018& 3&---&GES\\    
  NGC5904 M5  &I  &   -1.462  & 0.269 &31   & -1.36  &0.09  &0.552   &0.073 &80   & 0.317  &.031 &38&+0.31&He\\   
  NGC5927   & I &    ---    &  ---   &0   &  -0.29   & 0.07  & ---    & ---   & 0   &  ---   &---  & 0&-1.00&Di\\    
  NGC5946   & II &    -1.677 &  0.357& 2   &  -1.29   & 0.14  & 0.602  & ---  & 1   &  ---   &---  & 0&---&Kr\\    
  NGC5986   & II &    -1.786 &  0.194& 5   &  -1.63   & 0.08  & 0.649  & 0.103& 4   &  0.328 &---  & 1&+0.97&Kr\\    
  NGC6093 M80& II &    -2.111 &  0.194& 4   &  -1.75   & 0.08  & 0.688  & 0.075& 4   &  0.388 &0.060& 6&+0.93&Bl\\    
  NGC6101    &II  &   -2.179  & 0.168 &2    & -1.98    &0.07  &0.834   &0.080 &2    & 0.374  &.038 &13&+0.84&Sq\\  
  NGC6121 M4  &I  &   -1.237  & 0.161 &31   & -1.18    &0.02   &0.548   &0.078 &31   & 0.292  &.038 &14&-0.06&Kr\\   
  NGC6139   & II &    -1.616 &  ---  & 1   &  -1.56 & 0.02 & 0.730  & 0.030& 2   &  0.417 &---  & 1&---&Kr\\    
  NGC6144   & I &    ---    &  ---  & --   &  -1.82   & 0.04 & ---    & ---  & --   &  ---   &---  & 0&+1.00&Bl\\    
  NGC6171 M107& I &    -1.143 &  0.116& 14  &  -1.01 & 0.12 & 0.534  & 0.070& 13  &  0.292 &0.023& 6&-0.73&Bl\\    
  NGC6205 M13& II &    -2.076 &  0.274& 1   &  -1.58 & ---  & 0.750  & ---  & 1   &  0.360 &0.047& 7&+0.97&GES\\    
  NGC6218 M12& II &    ---    &  ---  & 0   &  -1.33   & 0.02  & ---    & ---  & 0   &  ---   &---  & 0&+0.97&Di\\    
  NGC6229    &I  &   -1.351  & 0.146 &21   & -1.32  &0.07  &0.557   &0.049 &36   & 0.328  &.033 &13&+0.24&He\\   
  NGC6235   & I &    -1.594 &  0.168& 2   &  -1.38   & 0.07  & 0.600  & 0.016& 2   &  0.352 &---  & 1&---&GES\\    
  NGC6254 M10& II &    ---    & ---  & 0   &  -1.52   & ---  & ---    & ---  & 0   &  ---   &---  & 0&+1.00&Kr\\    
  NGC6266 M62 &I  &   -1.208  & 0.226 &92   & -1.15  &0.11  &0.548   &0.069 &42   & 0.296  &.027 &28&+0.55&Bl\\   
  NGC6273 M19& I &    -1.603 &  0.270& 1   &  -1.76   & 0.07  & 0.507  & ---  & 1   &  ---   &---  & 0&---&Kr\\    
  NGC6284   & I &    -1.456 &  0.193& 4   &  -1.31   &0.09  & 0.559  & --- & 1   &  ---   &---  & 0&+0.88&Kr\\    
  NGC6287   & ? &    -1.871 &  0.255& 1   &  -2.12   & 0.09  & ---    & ---  & 0   &  ---   &---  & 0&--&Kr\\    
  NGC6293   & ? &    -1.922 &  0.291& 1   &  -2.01   & 0.14  & ---    & ---  &--   &  0.353 &0.016& 2&--&Bl\\    
  NGC6316   & I &    ---    &  ---  & --   &  -0.36  & 0.14  & ---    & ---  & --   &  ---   &---  & --&--&Kr\\    
  NGC6235   & I &    ---    &  ---  & --   &  -1.38   & 0.07  & 0.599  & 0.016& 2   &  0.352 &---  & 1&--&GES\\    
  NGC6333 M9 & II &    -1.772 &  0.070& 6   &  -1.81 & 0.13 & 0.624  & 0.029& 5   &  0.341 &0.024& 7&+0.87&Kr\\
 NGC6341 M92& II &    -2.010 &  0.101& 7   &  -2.14 & 0.18 & 0.646  & 0.036& 9   &  0.344 &0.042& 5&+0.91&GES\\    

\hline
\end{tabular}
\end{center}
\end{table*}

\begin{table*}
\addtocounter{table}{-1}
\caption{Continue}
\begin{center}
	\begin{tabular}{lccccccccrccrcc}
			\hline
  Name   & Oo Type &[Fe/H]$_{\rm Pr}$& s& N &[Fe/H]$^{\rm UVES}_{\rm AF}$ & s & <P$_{\rm ab}$> & s &  N & <P$_{\rm c}$> &s & N&$HBt$&Group$^*$\\    
            \hline
  
  NGC6352   & ? &    ---    &  ---  & --   &  -0.62   & 0.05  & ---    & ---  & --  &  ---   &---  & --&-1.00&Di\\    
  NGC6355   &II &    -1.577 &  0.317& 2   &  -1.33   & 0.14  & 0.702  & 0.102& 4   &  ---   &---  & --&---&Bl\\    
  NGC6356   & I &    ---    &  ---  & --   &  -0.35   & 0.14 & ---    & ---  & --   &  ---   &---  & --&---&Kr\\   
  NGC6362    &I  &   -1.249  & 0.216 &18   & -1.11  &0.06  &0.556   &0.063 &16   & 0.291  &.034 &13&-0.58&Di\\   
  NGC6366   & ? &    -0.795 &  0.254& 1   &  -0.77 & ---  & 0.513  & ---  & 1   &  ---   &---  & 0&-0.97&Di\\    
  NGC6380 Ter1 & I &    -0.756 &  0.535& 1   &  -0.40   & 0.09  & ---    & ---  & --   &  ---   &---  & 0&---&Bl\\    
  NGC6388   & III &    -1.906 &  0.136& 2   &  -0.93& 0.055& 0.763  & 0.051& 2   &  0.340 &0.100& 7&-1.00&Bl\\    
  NGC6401   &I &    -1.145 &  0.176& 19  &  -1.01 & 0.14 & 0.545  & 0.043& 15  &  0.303 &0.027& 5&---&Bl\\   
  NGC6402    &I  &   -1.395  & 0.266 &33   & -1.44  &0.17  &0.599   &0.096 &42   & 0.309  &.040 &47&+0.65&Kr\\   
  NGC6426    &II  &   -2.115  & 0.092 &8    & -2.15    &0.35   &0.703   &0.049 &6    & 0.347  &.040 &51&+0.58&He\\   
  NGC6441    &III &   -1.948  & 0.465 &3    & -1.02  &0.17  &0.754   &0.083 &15   & 0.376  &.073 &11&-0.73&Di\\   
  NGC6453   & II &    -2.170 &  0.315& 1   &  -1.48  & 0.14 & 0.733  & 0.003& 2   &  0.323 &---  & 1&---&Bl\\    
  NGC6496   & ? &    ---    &  ---  & --   &  -0.46   & 0.07  & ---    & ---  & --   &  ---   &---  & --&-1.00&Di\\    
  NGC6517   & II &    -2.063 &  0.271& 1   &  -1.24   & 0.14 & ---    & ---  & --  &  ---   &---  & --&---&Di\\    
  NGC6522   & I &    -1.656 &  0.466& 1   &  -1.16 & 0.09 & 0.603  & 0.033& 3   &  0.282 &0.008& 3&---&Bl\\    
  NGC6540   & I &    ---    &  ---  & --   &  ---   & ---  & 0.450  & ---  & 1   &  ---   &---  & --&--&Bl\\    
  NGC6541   & ? &    ---    &  ---  & --   &  -1.82  & 0.08  & ---    & --- & --   &  ---   &---  & 0&+1.00&Kr\\    
  NGC6544   & I &    -1.243 &  0.282& 1   &  -1.47   & 0.07  & 0.572  & ---  & 1   &  ---   &---  & 0&---&Kr\\    
  NGC6553   & I &    ---    &  ---  & --   &  -0.16   &0.06  & 0.489  & ---  & 1   &  ---   &---  & --&---&Kr\\    
  NGC6558   & ? &    -1.28  &  0.09 & --   &  -1.20 & 0.13 & 0.677  & 0.069& 3   &  0.313 &---  & 1&+0.70&Bl\\    
  NGC6569   & I &    -0.801 &  0.927& 1   &  -0.72   & 0.14  & 0.570  & 0.063& 5   &  0.300 &0.021& 9&---&Kr\\    
  NGC6584    &I  &   -1.488  & 0.220 &20   & -1.50    &0.09   &0.575   &0.067 &38   & 0.309  &.023 &15&-0.15&He\\   
  NGC6626 M28& I &    -1.407 &  0.116& 4   &  -1.46   & 0.09  & 0.567  & 0.059& 7   &  0.309 &0.015& 5&+0.43&Bl\\    
  NGC6638   & I &    -1.558 &  ---  & 1   &  -0.90   & 0.07  & 0.610  & 0.066& 4   &  0.300 &0.021& 7&---&Bl\\    
  NGC6642   & I &    -1.062 &  0.077& 4   &  -1.19   & 0.14  & 0.447  & 0.011& 2   &  0.299 &0.008& 2&-0.04&Bl\\    
  NGC6656 M22 &II  &   -2.353  & 0.794 &1    & -1.70    &0.08   &0.625   &0.106 &11   & 0.334  &.035 &15&+0.91&GES\\   
  NGC6681 M70& ? &    ---    &  ---  & --   &  -1.62   & 0.08  & 0.564  & ---  & 1   &  0.372 &0.030& 2&+0.96&Kr\\    
  NGC6712   & I &    -1.341 &  0.135& 8   &  -1.04 & 0.06 & 0.578  & 0.058& 10  &  0.338 &0.058& 4&-0.62&Kr\\    
  NGC6715 M54 &I  &   -1.478  & 0.211 &41   & -1.44    &0.07  &0.598   &0.065 &96   & 0.331  &.037 &21&+0.54&Sg\\   
  NGC6717 Pal9& ? &    -1.250 &  0.825& 1   &  -1.26   & 0.07  & 0.575  & ---  & 1   &  ---   &---  & --&+0.98&Bl\\    
  NGC6723   & I &    -1.284 &  0.300& 8   &  -1.10   & 0.07  & 0.543  & 0.064& 30  &  0.292 &0.028& 7&-0.08&Bl\\    
  NGC6749   & II &    -2.155 &  0.402& 1   &  -1.61   & 0.09  & ---    & ---  & --   &  ---   &---  & --&--&Kr\\    
  NGC6752   & ? &    ---    &  ---  & --  &  -1.55   & 0.01  & ---    & ---  & --   &  ---   &---  & --&+1.00&Kr\\    
  NGC6760   & I &    ---    &  ---  & --   &  -0.40   & 0.14  & ---    & ---  & --   &  ---   &---  & --&--&Kr\\    
  NGC6779 M56& ? &    -2.239 &  0.270& 1   &  -2.00 & 0.09  & 0.906  & ---  & 1   &  0.358 &0.063& 3&+0.98&GES\\    
  NGC6809 M55& II &    -1.863 &  0.052& 2   &  -1.93 & 0.02 & 0.662  & 0.052& 4   &  0.355 &0.035& 8&+0.87&Kr\\    
  NGC6838 M71& I &    ---    &  ---  & --   &  -0.82   & 0.02 & ---    & ---  & --  &  ---   &---  & --&-1.00&Di\\    
  NGC6864 M75& I &    -1.385 &  0.281& 7   &  -1.29   & 0.14  & 0.586  & 0.080& 20  &  0.309 &0.050& 9&-0.07&GES\\    
  NGC6934   & I &    -1.526 &  0.215& 52  &  -1.45 & 0.14 & 0.573  & 0.063& 58  &  0.398 &0.040& 9&+0.25&Ug\\    
  NGC6981 M72& I &    -1.572 &  0.235& 30  &  -1.50 & 0.11 & 0.569  & 0.052& 34  &  0.312 &0.027& 6&+0.14&He\\    
  NGC7006   & I &    -1.529 &  0.240& 32  &  -1.42 & 0.13 & 0.564  & 0.051& 48  &  0.330 &0.038& 5&-0.28&Sq\\    
  NGC7078 M15 &II  &   -2.006  & 0.254 &37   & -2.35  &0.19  &0.642   &0.062 &46   & 0.363  &.043 &45&+0.67&GES\\   
  NGC7089 M2  &II  &   -1.840  & 0.195 &14   & -1.62  &0.05  &0.683   &0.084 &17   & 0.324  &.036 &12&+0.96&GES\\   
  NGC7099 M30& II &    -2.079 &  0.172& 4   &  -2.18   & 0.05  & 0.693  & 0.033& 4   &  0.337 &0.011& 2&+0.89&GES\\    
  NGC7492   & II &    -1.911 &  0.312& 1   &  -1.93 & ---  & 0.805  & ---  & 1   &  0.412 &---  & 1&+0.76&He\\    
  Pal2      & I &    -1.198 &  0.209& 9   &  -1.09 & 0.55 & 0.580  & 0.053& 14  &  0.373 &---  & 1&---&Ug\\ 
  Pal3      & I &    -1.146 &  0.830& 1   &  -1.67   & 0.08 & ---    & ---  & --   &  ---   &---  & --&---&Sq\\    
  Pal5      & I &    ---    &  ---  & --   &  -2.10   & 0.08  & ---    & ---  & --  &  0.299 &0.030& 5&---&He\\    
  Pal10     & I &    ---    &  ---  & --  &  ---   & ---  & ---    & ---  & --  &  ---   &---  & --&---&Di\\    
  Pal13     & I &    -1.162 &  0.248& 3   &  -1.56 & 0.15 & 0.571  & 0.025& 3   &  ---   &---  & --&-0.30&Sq\\    
  Pal15     & II &    ---    &  ---  & --   &  ---   & ---  & ---    & ---  & --   &  ---   &---  & --&---&Sq\\    
  Rup106   & I &    -1.660 &  0.109& 10  &  -1.78   & 0.08  & 0.620  & 0.022& 9   &  ---   &---  & --&-0.82&He\\    
  Ryu059   & I &    -2.451 &  0.393& 1   &  ---   & ---  & ---    & ---  & --   &  ---   &---  & --&---&---\\    
  Ter1 HP2   & I &    -1.076 &  0.390& 1   &  -1.29   & 0.09  & 0.593  & 0.073& 8   &  0.306 &0.011& 3&---&Bl\\    
  Ter3      & I &    ---    &  ---  & --   &  ---   & ---  & ---    & ---  & --   &  ---   &---  & --&---&---\\    
  Ter5      & II &    ---    &  ---  & --   &  +0.16   & 0.09  & 0.680  & 0.040& 2   &  ---   &---  & --&---&---\\    
  Ter8      & II &    -2.106 &  0.344 & 1   &  ---   & ---  & 0.686  & ---  & 1   &  0.392 &---  & 1&1.00&Sg\\    
  Ter10     & I &    ---    &  ---  & --   &  ---   & ---  & 0.691  & 0.043& 4   &  ---   &---  & --&---&Kr\\    
  Ter12     & II &    ---    &  ---  & --   &  ---   & ---  & ---    & ---  & --  &  ---   &---  & --&---&Kr\\   

\hline
\end{tabular} 
\end{center}
\quad $*$ Groups code: Bl-Bulge, Di-Disc, GES-Gaia-Enceladus-Sausage, He-Helmi, Kr-Kraken, Sg-Sagittarius, Sq-Sequoia, Ug-ungrouped.
\end{table*}

\section{Results: The electronic tables} \label{results}

In this study, we provide a membership probability for the catalog of variable stars, CVSGC, \citep{Clement2001}, associated with the MW globular clusters. To this end, we used the results of the analysis performed by \citet{Vasiliev2021} using the \textit{Gaia} astrometric data. We matched the CVSGC catalog with stars analyzed by \citet{Vasiliev2021} using available equatorial coordinates to obtain individual membership probabilities. Stars that were outside the analysis footprint of \citet{Vasiliev2021} or variables that were not analyzed by \citet{Vasiliev2021} we assigned a negative membership probability. The cases with the highest difference between the number of detected variable stars toward a certain cluster versus the variable content (based on membership probabilities) are mostly due to their location toward or within the Galactic bulge. Therefore a large foreground and background population of variable stars is present. The newly obtained information on memberships of variables stars to a given globular clusters will be included in the new version of the CVSGC\footnote{Available here: \url{https://www.astro.utoronto.ca/~cclement/cat/listngc.html}}.

We have generated tables listing the variable stars presently reported in the CVSGC and their membership probabilities in 121 globular clusters listed in Table \ref{FevsPab}. These tables shall be available in the Centre de Donnés astronomiques de Strasbourg database (CDS). Table \ref{N5053_ex} for NGC 5053 is included here as an example.

\begin{table*}
\caption{Example of one of the electronic tables for the variable stars in NGC 5053 and their cluster membership probabilities.}
\begin{center}
\label{N5053_ex}
	\begin{tabular}{cccccccc}
			\hline
  Variable& Coo. frame$^*$ &  ra (deg) & dec (deg) & Var type & Period (day) & Mem. prob. &Prob. name\\
			\hline 
1 &R0 &198.99721 &17.74097 &RR0 &0.6472 &0.999 &m1\\
2 &R0 &199.05154 &17.69628 &RR1 &0.379 &1.0 &m1\\
3 &R0 &199.14842 &17.73775 &RR0 &0.5929 &1.0 &m1\\
4 &R0 &199.11679 &17.66533 &RR0 &0.6671 &1.0 &m1\\
5 &R0 &199.17175 &17.63631 &RR0 &0.7149 &1.0 &m1\\
6 &R0 &199.14425 &17.71867 &RR1 &0.2922 &0.999 &m1\\
7 &R0 &199.08213 &17.74398 &RR1 &0.3519 &1.0 &m1\\
8 &R0 &199.14179 &17.71120 &RR1 &0.3628 &1.0 &m1\\
9 &R0 &199.04954 &17.80300 &RR0 &0.7402 &0.999 &m1\\
10 &R0 &199.13554 &17.71289 &RR0 &0.7759 &0.999 &m1\\
11 &R0 &199.13433 &17.70273 &SXPHE &0.0368 &0.999 &m1\\
12 &R0 &199.10546 &17.73009 &SXPHE &0.035 &0.999 &m1\\
13 &R0 &199.09937 &17.70031 &SXPHE &0.0342 &0.999 &m1\\
14 &R0 &199.07437 &17.68131 &SXPHE &0.0393 &0.999 &m1\\
15 &R0 &199.07171 &17.67486 &SXPHE &0.0356 &0.999 &m1\\ 
\hline
\end{tabular}
\end{center}
\quad $*$
R0= J2000 coordinates; R5= J1950 coordinates
\end{table*}

Using the newly obtained cleaned variables stars datasets for MW globular clusters we revisit the analysis done by \citet{Catelan2009} on the Oosterhoff dichotomy. We recalculate average pulsation periods for fundamental-mode RR~Lyrae stars associated with individual globular clusters. Using photometric metallicities derived based on \textit{Gaia} light curves and based on our analysis we revisit their Oosterhoff classification and reproduce the results outlined in \citet{Catelan2009}. Only a handful of globular clusters were reclassified, mostly moving from the Oosterhoff intermediate to the Oosterhoff type I group. The nearly depleted Oosterhoff gap separating both Oosterhoff types is partially occupied by Globular clusters associated with MW dwarf galaxies and with clusters with a low number of fundamental pulsators.

\noindent
{\bf Acknowledgements.} AAF is grateful to the  European Southern Observatory (ESO) for warm hospitality during the preparation of this work. We thank Dr. Mario Yepez for comments and technical assistance with some of the figures. AAF also thankfully acknowledges the sabbatical support granted by the program PASPA of the DGAPA-UNAM.  We have been benefited from the support of DGAPA-UNAM through projects IG100620 and IN103024.

{\bf Data Availability:} The data underlying this article shall be available in electronic form in the Centre de Donnés astronomiques de Strasbourg database (CDS), and can also be shared on request to the corresponding author.

\bibliographystyle{mnras}
\bibliography{CVSGC} 

\begin{thebibliography}{}
\makeatletter
\relax
\def\mn@urlcharsother{\let\do\@makeother \do\$\do\&\do\#\do\^\do\_\do\%\do\~}
\def\mn@doi{\begingroup\mn@urlcharsother \@ifnextchar [ {\mn@doi@} {\mn@doi@[]}}
\def\mn@doi@[#1]#2{\def\@tempa{#1}\ifx\@tempa\@empty \href {http://dx.doi.org/#2} {doi:#2}\else \href {http://dx.doi.org/#2} {#1}\fi \endgroup}
\def\mn@eprint#1#2{\mn@eprint@#1:#2::\@nil}
\def\mn@eprint@arXiv#1{\href {http://arxiv.org/abs/#1} {{\tt arXiv:#1}}}
\def\mn@eprint@dblp#1{\href {http://dblp.uni-trier.de/rec/bibtex/#1.xml} {dblp:#1}}
\def\mn@eprint@#1:#2:#3:#4\@nil{\def\@tempa {#1}\def\@tempb {#2}\def\@tempc {#3}\ifx \@tempc \@empty \let \@tempc \@tempb \let \@tempb \@tempa \fi \ifx \@tempb \@empty \def\@tempb {arXiv}\fi \@ifundefined {mn@eprint@\@tempb}{\@tempb:\@tempc}{\expandafter \expandafter \csname mn@eprint@\@tempb\endcsname \expandafter{\@tempc}}}

\bibitem[\protect\citeauthoryear{{Arellano Ferro}}{{Arellano Ferro}}{2024}]{Arellano2024}
{Arellano Ferro} A.,  2024, \mn@doi [IAU Symposium] {10.1017/S1743921323002880}, \href {https://ui.adsabs.harvard.edu/abs/2024IAUS..376..222A} {376, 222 (AF24)}

\bibitem[\protect\citeauthoryear{{Arellano Ferro}, {Zerpa Guillen}, {Yepez}, {Bustos Fierro}, {Prudil}  \& {P{\'e}rez Parra}}{{Arellano Ferro} et~al.}{2024}]{Arellano2024b}
{Arellano Ferro} A.,  {Zerpa Guillen} L.~J.,  {Yepez} M.~A.,  {Bustos Fierro} I.~H.,  {Prudil} Z.,   {P{\'e}rez Parra} C.~E.,  2024, \mn@doi [\mnras] {10.1093/mnras/stae1609}, \href {https://ui.adsabs.harvard.edu/abs/2024MNRAS.tmp.1572A} {}

\bibitem[\protect\citeauthoryear{{Baade}}{{Baade}}{1956}]{Baade1956}
{Baade} W.,  1956, \mn@doi [\pasp] {10.1086/126870}, \href {https://ui.adsabs.harvard.edu/abs/1956PASP...68....5B} {68, 5}

\bibitem[\protect\citeauthoryear{{Bla{\v z}ko}}{{Bla{\v z}ko}}{1907}]{Blazhko1907}
{Bla{\v z}ko} S.,  1907, \mn@doi [Astronomische Nachrichten] {10.1002/asna.19071752002}, \href {http://adsabs.harvard.edu/abs/1907AN....175..325B} {175, 325}

\bibitem[\protect\citeauthoryear{{Bono}, {Caputo}  \& {Stellingwerf}}{{Bono} et~al.}{1994}]{Bono1994}
{Bono} G.,  {Caputo} F.,   {Stellingwerf} R.~F.,  1994, \mn@doi [\apj] {10.1086/173806}, \href {https://ui.adsabs.harvard.edu/abs/1994ApJ...423..294B} {423, 294}

\bibitem[\protect\citeauthoryear{{Cacciari}, {Corwin}  \& {Carney}}{{Cacciari} et~al.}{2005}]{Cacciari2005}
{Cacciari} C.,  {Corwin} T.~M.,   {Carney} B.~W.,  2005, \aj, \href {http://adsabs.harvard.edu/abs/2005AJ....129..267C} {129, 267}

\bibitem[\protect\citeauthoryear{{Callingham}, {Cautun}, {Deason}, {Frenk}, {Grand}  \& {Marinacci}}{{Callingham} et~al.}{2022}]{Callingham2022}
{Callingham} T.~M.,  {Cautun} M.,  {Deason} A.~J.,  {Frenk} C.~S.,  {Grand} R. J.~J.,   {Marinacci} F.,  2022, \mn@doi [\mnras] {10.1093/mnras/stac1145}, \href {https://ui.adsabs.harvard.edu/abs/2022MNRAS.513.4107C} {513, 4107}

\bibitem[\protect\citeauthoryear{{Caputo}, {Castellani}  \& {Tornambe}}{{Caputo} et~al.}{1978}]{Caputo1978}
{Caputo} F.,  {Castellani} V.,   {Tornambe} A.,  1978, \aap, \href {https://ui.adsabs.harvard.edu/abs/1978A&A....67..107C} {67, 107}

\bibitem[\protect\citeauthoryear{{Carretta} \& {Bragaglia}}{{Carretta} \& {Bragaglia}}{2022}]{Carretta2022}
{Carretta} E.,  {Bragaglia} A.,  2022, \mn@doi [\aap] {10.1051/0004-6361/202142563}, \href {https://ui.adsabs.harvard.edu/abs/2022A&A...659A.122C} {659, A122}

\bibitem[\protect\citeauthoryear{{Carretta}, {Bragaglia}, {Gratton}, {D'Orazi}  \& {Lucatello}}{{Carretta} et~al.}{2009}]{Carretta2009}
{Carretta} E.,  {Bragaglia} A.,  {Gratton} R.,  {D'Orazi} V.,   {Lucatello} S.,  2009, \aap, \href {http://adsabs.harvard.edu/abs/2009A%26A...508..695C} {508, 695}

\bibitem[\protect\citeauthoryear{{Catelan}}{{Catelan}}{2009}]{Catelan2009}
{Catelan} M.,  2009, \mn@doi [\apss] {10.1007/s10509-009-9987-8}, \href {https://ui.adsabs.harvard.edu/abs/2009Ap&SS.320..261C} {320, 261}

\bibitem[\protect\citeauthoryear{{Chadid}, {Sneden}  \& {Preston}}{{Chadid} et~al.}{2017}]{Chadid2017}
{Chadid} M.,  {Sneden} C.,   {Preston} G.~W.,  2017, \mn@doi [\apj] {10.3847/1538-4357/835/2/187}, \href {https://ui.adsabs.harvard.edu/abs/2017ApJ...835..187C} {835, 187}

\bibitem[\protect\citeauthoryear{{Clement} et~al.,}{{Clement} et~al.}{2001}]{Clement2001}
{Clement} C.~M.,  et~al., 2001, \aj, \href {http://adsabs.harvard.edu/abs/2001AJ....122.2587C} {122, 2587}

\bibitem[\protect\citeauthoryear{{Clementini}, {Gratton}, {Bragaglia}, {Ripepi}, {Martinez Fiorenzano}, {Held}  \& {Carretta}}{{Clementini} et~al.}{2005}]{Clementini2005}
{Clementini} G.,  {Gratton} R.~G.,  {Bragaglia} A.,  {Ripepi} V.,  {Martinez Fiorenzano} A.~F.,  {Held} E.~V.,   {Carretta} E.,  2005, \mn@doi [\apjl] {10.1086/491789}, \href {https://ui.adsabs.harvard.edu/abs/2005ApJ...630L.145C} {630, L145}

\bibitem[\protect\citeauthoryear{{Clementini} et~al.,}{{Clementini} et~al.}{2023}]{Clementini2023}
{Clementini} G.,  et~al., 2023, \mn@doi [\aap] {10.1051/0004-6361/202243964}, \href {https://ui.adsabs.harvard.edu/abs/2023A&A...674A..18C} {674, A18}

\bibitem[\protect\citeauthoryear{{Crestani} et~al.,}{{Crestani} et~al.}{2021}]{Crestani2021}
{Crestani} J.,  et~al., 2021, \mn@doi [\apj] {10.3847/1538-4357/abd183}, \href {https://ui.adsabs.harvard.edu/abs/2021ApJ...908...20C} {908, 20}

\bibitem[\protect\citeauthoryear{{D{\'e}k{\'a}ny}, {Grebel}  \& {Pojma{\'n}ski}}{{D{\'e}k{\'a}ny} et~al.}{2021}]{Dekany2021}
{D{\'e}k{\'a}ny} I.,  {Grebel} E.~K.,   {Pojma{\'n}ski} G.,  2021, \mn@doi [\apj] {10.3847/1538-4357/ac106f}, \href {https://ui.adsabs.harvard.edu/abs/2021ApJ...920...33D} {920, 33}

\bibitem[\protect\citeauthoryear{{Fabrizio} et~al.,}{{Fabrizio} et~al.}{2019}]{Fabrizio2019}
{Fabrizio} M.,  et~al., 2019, \mn@doi [\apj] {10.3847/1538-4357/ab3977}, \href {https://ui.adsabs.harvard.edu/abs/2019ApJ...882..169F} {882, 169}

\bibitem[\protect\citeauthoryear{{Fabrizio} et~al.,}{{Fabrizio} et~al.}{2021}]{Fabrizio2021}
{Fabrizio} M.,  et~al., 2021, \mn@doi [\apj] {10.3847/1538-4357/ac1115}, \href {https://ui.adsabs.harvard.edu/abs/2021ApJ...919..118F} {919, 118}

\bibitem[\protect\citeauthoryear{{For}, {Sneden}  \& {Preston}}{{For} et~al.}{2011}]{For2011}
{For} B.-Q.,  {Sneden} C.,   {Preston} G.~W.,  2011, \mn@doi [\apjs] {10.1088/0067-0049/197/2/29}, \href {https://ui.adsabs.harvard.edu/abs/2011ApJS..197...29F} {197, 29}

\bibitem[\protect\citeauthoryear{{Horta} et~al.,}{{Horta} et~al.}{2021}]{Horta2021}
{Horta} D.,  et~al., 2021, \mn@doi [\mnras] {10.1093/mnras/staa2987}, \href {https://ui.adsabs.harvard.edu/abs/2021MNRAS.500.1385H} {500, 1385}

\bibitem[\protect\citeauthoryear{{Jurcsik} \& {Kov{\'a}cs}}{{Jurcsik} \& {Kov{\'a}cs}}{1996}]{Jurcsik1996}
{Jurcsik} J.,  {Kov{\'a}cs} G.,  1996, \aap, \href {http://adsabs.harvard.edu/abs/1996A%26A...312..111J} {312, 111}

\bibitem[\protect\citeauthoryear{{Kirby}, {Cohen}, {Guhathakurta}, {Cheng}, {Bullock}  \& {Gallazzi}}{{Kirby} et~al.}{2013}]{Kirby2013}
{Kirby} E.~N.,  {Cohen} J.~G.,  {Guhathakurta} P.,  {Cheng} L.,  {Bullock} J.~S.,   {Gallazzi} A.,  2013, \mn@doi [\apj] {10.1088/0004-637X/779/2/102}, \href {https://ui.adsabs.harvard.edu/abs/2013ApJ...779..102K} {779, 102}

\bibitem[\protect\citeauthoryear{{Koppelman}, {Helmi}, {Massari}, {Roelenga}  \& {Bastian}}{{Koppelman} et~al.}{2019}]{Koppelman2019}
{Koppelman} H.~H.,  {Helmi} A.,  {Massari} D.,  {Roelenga} S.,   {Bastian} U.,  2019, \mn@doi [\aap] {10.1051/0004-6361/201834769}, \href {https://ui.adsabs.harvard.edu/abs/2019A&A...625A...5K} {625, A5}

\bibitem[\protect\citeauthoryear{{Kruijssen} et~al.,}{{Kruijssen} et~al.}{2020}]{Kruijssen2020}
{Kruijssen} J.~M.~D.,  et~al., 2020, \mn@doi [\mnras] {10.1093/mnras/staa2452}, \href {https://ui.adsabs.harvard.edu/abs/2020MNRAS.498.2472K} {498, 2472}

\bibitem[\protect\citeauthoryear{{Kunder}, {Stetson}, {Catelan}, {Walker}  \& {Amigo}}{{Kunder} et~al.}{2013a}]{Kunder2013a}
{Kunder} A.,  {Stetson} P.~B.,  {Catelan} M.,  {Walker} A.~R.,   {Amigo} P.,  2013a, \aj, \href {http://adsabs.harvard.edu/abs/2013AJ....145...33K} {145, 33}

\bibitem[\protect\citeauthoryear{{Kunder} et~al.,}{{Kunder} et~al.}{2013b}]{Kunder2013b}
{Kunder} A.,  et~al., 2013b, \aj, \href {http://adsabs.harvard.edu/abs/2013AJ....146..119K} {146, 119}

\bibitem[\protect\citeauthoryear{{Kunder}, {Prudil}, {Skaggs}, {Reggiani}, {Nataf}, {Hughes}, {Covey}  \& {Devine}}{{Kunder} et~al.}{2024}]{Kunder2024}
{Kunder} A.,  {Prudil} Z.,  {Skaggs} C.,  {Reggiani} H.,  {Nataf} D.~M.,  {Hughes} J.,  {Covey} K.~R.,   {Devine} K.,  2024, \mn@doi [arXiv e-prints] {10.48550/arXiv.2407.01515}, \href {https://ui.adsabs.harvard.edu/abs/2024arXiv240701515K} {p. arXiv:2407.01515}

\bibitem[\protect\citeauthoryear{{Lane}, {Bovy}  \& {Mackereth}}{{Lane} et~al.}{2023}]{Lane2023}
{Lane} J. M.~M.,  {Bovy} J.,   {Mackereth} J.~T.,  2023, \mn@doi [\mnras] {10.1093/mnras/stad2834}, \href {https://ui.adsabs.harvard.edu/abs/2023MNRAS.526.1209L} {526, 1209}

\bibitem[\protect\citeauthoryear{{Lee}, {Demarque}  \& {Zinn}}{{Lee} et~al.}{1994}]{Lee1994}
{Lee} Y.-W.,  {Demarque} P.,   {Zinn} R.,  1994, \mn@doi [\apj] {10.1086/173803}, \href {https://ui.adsabs.harvard.edu/abs/1994ApJ...423..248L} {423, 248}

\bibitem[\protect\citeauthoryear{{Lindegren} et~al.,}{{Lindegren} et~al.}{2021}]{Lindegren2021}
{Lindegren} L.,  et~al., 2021, \mn@doi [\aap] {10.1051/0004-6361/202039709}, \href {https://ui.adsabs.harvard.edu/abs/2021A&A...649A...2L} {649, A2}

\bibitem[\protect\citeauthoryear{{Luongo}, {Ripepi}, {Marconi}, {Prudil}, {Rejkuba}, {Clementini}  \& {Longo}}{{Luongo} et~al.}{2024}]{Luongo2024}
{Luongo} E.,  {Ripepi} V.,  {Marconi} M.,  {Prudil} Z.,  {Rejkuba} M.,  {Clementini} G.,   {Longo} G.,  2024, \mn@doi [arXiv e-prints] {10.48550/arXiv.2409.04259}, \href {https://ui.adsabs.harvard.edu/abs/2024arXiv240904259L} {p. arXiv:2409.04259}

\bibitem[\protect\citeauthoryear{{Massari}, {Koppelman}  \& {Helmi}}{{Massari} et~al.}{2019}]{Massari2019}
{Massari} D.,  {Koppelman} H.~H.,   {Helmi} A.,  2019, \mn@doi [\aap] {10.1051/0004-6361/201936135}, \href {https://ui.adsabs.harvard.edu/abs/2019A&A...630L...4M} {630, L4}

\bibitem[\protect\citeauthoryear{{Oosterhoff}}{{Oosterhoff}}{1939}]{Oosterhoff1939}
{Oosterhoff} P.~T.,  1939, The Observatory, \href {https://ui.adsabs.harvard.edu/abs/1939Obs....62..104O} {62, 104}

\bibitem[\protect\citeauthoryear{{Oosterhoff}}{{Oosterhoff}}{1944}]{Oosterhoff1944}
{Oosterhoff} P.~T.,  1944, \bain, \href {https://ui.adsabs.harvard.edu/abs/1944BAN....10...55O} {10, 55}

\bibitem[\protect\citeauthoryear{{Sandage}}{{Sandage}}{1970}]{Sandage1970}
{Sandage} A.,  1970, \mn@doi [\apj] {10.1086/150715}, \href {https://ui.adsabs.harvard.edu/abs/1970ApJ...162..841S} {162, 841}

\bibitem[\protect\citeauthoryear{{Sandage}}{{Sandage}}{1981}]{Sandage1981a}
{Sandage} A.,  1981, \mn@doi [\apj] {10.1086/159140}, \href {https://ui.adsabs.harvard.edu/abs/1981ApJ...248..161S} {248, 161}

\bibitem[\protect\citeauthoryear{{Sandage}, {Katem}  \& {Sandage}}{{Sandage} et~al.}{1981}]{Sandage1981b}
{Sandage} A.,  {Katem} B.,   {Sandage} M.,  1981, \mn@doi [\apjs] {10.1086/190734}, \href {https://ui.adsabs.harvard.edu/abs/1981ApJS...46...41S} {46, 41}

\bibitem[\protect\citeauthoryear{{Shapley}}{{Shapley}}{1917}]{Shapley1917}
{Shapley} H.,  1917, \mn@doi [Proceedings of the National Academy of Science] {10.1073/pnas.3.7.479}, \href {https://ui.adsabs.harvard.edu/abs/1917PNAS....3..479S} {3, 479}

\bibitem[\protect\citeauthoryear{{Shapley}}{{Shapley}}{1918}]{Shapley1918}
{Shapley} H.,  1918, \mn@doi [\apj] {10.1086/142423}, \href {https://ui.adsabs.harvard.edu/abs/1918ApJ....48..154S} {48, 154}

\bibitem[\protect\citeauthoryear{{Sneden}, {Preston}, {Chadid}  \& {Adam{\'o}w}}{{Sneden} et~al.}{2017}]{Sneden2017}
{Sneden} C.,  {Preston} G.~W.,  {Chadid} M.,   {Adam{\'o}w} M.,  2017, \mn@doi [\apj] {10.3847/1538-4357/aa8b10}, \href {https://ui.adsabs.harvard.edu/abs/2017ApJ...848...68S} {848, 68}

\bibitem[\protect\citeauthoryear{{Torelli} et~al.,}{{Torelli} et~al.}{2019}]{Torelli2019}
{Torelli} M.,  et~al., 2019, \mn@doi [\aap] {10.1051/0004-6361/201935995}, \href {https://ui.adsabs.harvard.edu/abs/2019A&A...629A..53T} {629, A53}

\bibitem[\protect\citeauthoryear{{Vasiliev} \& {Baumgardt}}{{Vasiliev} \& {Baumgardt}}{2021}]{Vasiliev2021}
{Vasiliev} E.,  {Baumgardt} H.,  2021, \mn@doi [\mnras] {10.1093/mnras/stab1475}, \href {https://ui.adsabs.harvard.edu/abs/2021MNRAS.505.5978V} {505, 5978}

\bibitem[\protect\citeauthoryear{{Yepez}, {Arellano Ferro}  \& {Deras}}{{Yepez} et~al.}{2020}]{Yepez2020}
{Yepez} M.~A.,  {Arellano Ferro} A.,   {Deras} D.,  2020, \mn@doi [\mnras] {10.1093/mnras/staa637}, \href {https://ui.adsabs.harvard.edu/abs/2020MNRAS.494.3212Y} {494, 3212}

\bibitem[\protect\citeauthoryear{{Yepez}, {Arellano Ferro}, {Deras}, {Bustos Fierro}, {Muneer}  \& {Schr{\"o}der}}{{Yepez} et~al.}{2022}]{Yepez2022}
{Yepez} M.~A.,  {Arellano Ferro} A.,  {Deras} D.,  {Bustos Fierro} I.,  {Muneer} S.,   {Schr{\"o}der} K.~P.,  2022, \mn@doi [\mnras] {10.1093/mnras/stac054}, \href {https://ui.adsabs.harvard.edu/abs/2022MNRAS.511.1285Y} {511, 1285}

\bibitem[\protect\citeauthoryear{{Zinn} \& {West}}{{Zinn} \& {West}}{1984}]{Zinn1984}
{Zinn} R.,  {West} M.~J.,  1984, \apjs, \href {http://adsabs.harvard.edu/abs/1984ApJS...55...45Z} {55, 45}

\bibitem[\protect\citeauthoryear{{van Albada} \& {Baker}}{{van Albada} \& {Baker}}{1971}]{vanAlbada1971}
{van Albada} T.~S.,  {Baker} N.,  1971, \mn@doi [\apj] {10.1086/151144}, \href {https://ui.adsabs.harvard.edu/abs/1971ApJ...169..311V} {169, 311}

\bibitem[\protect\citeauthoryear{{van Albada} \& {Baker}}{{van Albada} \& {Baker}}{1973}]{vanAlbada1973}
{van Albada} T.~S.,  {Baker} N.,  1973, \mn@doi [\apj] {10.1086/152434}, \href {https://ui.adsabs.harvard.edu/abs/1973ApJ...185..477V} {185, 477}

\makeatother
\end{thebibliography}

\bsp	
\label{lastpage}

\end{document}